\documentclass[twocolumn,pra,superscriptaddress]{revtex4}
\usepackage{amsfonts}
\usepackage{amssymb}
\usepackage{amsmath}
\usepackage{epsfig}
\usepackage{color}
\usepackage{graphics, graphicx}
\usepackage{bbold}
\usepackage{psfrag}
\usepackage{mathcomp}
\usepackage{subfigure}
\usepackage{verbatim}
\usepackage[colorlinks,citecolor=blue]{hyperref}

\makeatletter

\newcommand{\Rmnum}[1]{\expandafter\@slowromancap\romannumeral #1@}
\makeatother

\begin{document}

\title{Topology-dependent quantum dynamics and entanglement-dependent topological pumping
       in superconducting qubit chains}

\author{Feng Mei}
\email{meifeng@sxu.edu.cn}
\affiliation{State Key Laboratory of Quantum Optics and Quantum Optics Devices, Institute
of Laser Spectroscopy, Shanxi University, Taiyuan, Shanxi 030006, China}
\affiliation{Collaborative Innovation Center of Extreme Optics, Shanxi
University,Taiyuan, Shanxi 030006, China}
\author{Gang Chen}
\email{chengang971@163.com}
\affiliation{State Key Laboratory of Quantum Optics and Quantum Optics Devices, Institute
of Laser Spectroscopy, Shanxi University, Taiyuan, Shanxi 030006, China}
\affiliation{Collaborative Innovation Center of Extreme Optics, Shanxi
University,Taiyuan, Shanxi 030006, China}
\author{Lin Tian}
\email{ltian@ucmerced.edu}
\affiliation{School of Natural Sciences, University of California, Merced, California
95343, USA}
\author{Shi-Liang Zhu}
\email{slzhu@nju.edu.cn}
\affiliation{National Laboratory of Solid State Microstructures, School of Physics,
Nanjing University, Nanjing 210093, China}
\affiliation{Guangdong Provincial Key Laboratory of Quantum Engineering and Quantum
Materials, SPTE, South China Normal University, Guangzhou 510006, China}

\author{Suotang Jia}
\affiliation{State Key Laboratory of Quantum Optics and Quantum Optics Devices, Institute
of Laser Spectroscopy, Shanxi University, Taiyuan, Shanxi 030006, China}
\affiliation{Collaborative Innovation Center of Extreme Optics, Shanxi
University,Taiyuan, Shanxi 030006, China}

\date{\today }

\begin{abstract}
 We propose a protocol using a tunable Xmon qubit chain to construct generalized Su-Schrieffer-Heeger (SSH) models that support various topological phases. We study the time evolution of a single-excitation quantum state in a SSH-type qubit chain and find that such dynamics is linked to topological winding number. We also investigate the adiabatic transfer of a single-excitation quantum state in a generalized SSH-type qubit chain and show that this process can be connected with topological Chern number and be used to generate a novel entanglement-dependent topological pumping. All results have been demonstrated to be robust against qubit coupling imperfections and can be observed in a short Xmon qubit chain. Our study provides a simple method to directly measure topological invariants rooted in momentum space using quantum dynamics in real space.
\end{abstract}

\maketitle

\section{Introduction}
Supercounducting circuits nowadays have been widely recognized as one of the leading quantum systems for quantum computation \cite{Devoret2013,YouNori,MartinisNPJ,Gambetta2015}. The fundamental challenge in building a full-fledged superconducting quantum computer is to balance high coherence and straightforward connectivity. Remarkable experimental progresses have recently been made in this regard \cite{Martinis2013,Martinis2014,Surfacecode,Chow2014,Corcoles2015,Takita2016,Brecht2016,Underwood2016}. In particular, Xmon qubits have been shown to possess excellent scalability simultaneously with high coherence \cite{Martinis2013}. Meanwhile, the coupling between Xmon qubits can be dynamically varied through a g-mon coupler \cite{Martinis2014}. Moreover, it is now believed that superconducting circuit can be further scaled up to several tens of qubits and show its quantum supremacy in the near term \cite{Boixo2016}. Such state-of-the-art enables superconducting qubit chains to be a promising platform for implementing large-scale quantum simulation \cite{NoriQS2009,NoriQS2014,CircuitQS,LeHur,PolaritonQS,Rosario2010,Rosario2007,Solano2014,Solano2015,
Juan2008,Juan2014,Tian2010,Tian2013,Tian2014,Zoller,Marquardt,Ashhab2014,Kapit,Sun,Chen,Zhou,
Koch,Ciuti,Schmidt,Sanders,Wang,Barends2015,Wallraff2015,Siddiqi2015,Roushan2016,Malley2016,Houck2017}.

On the other hand, searching topological states in cold atoms as well as photonic systems has recently become a rapidly growing research field \cite{CATopo,PHTopo1,PHTopo2}. In the context of superconducting circuits, some topological states and effects have also been theoretically studied \cite{You,Girvin,Taylor,KapitPRX,Tsomok2010,Gaspar2011,Abdum2013,Mei2015,Mei2016,Hu2016a,Hu2016b,Angelakis,Liu,Hur2018}. Experimentally, several progresses studying topological phenomenons recently have been made in superconducting qubits and resonators \cite{Schroer2014,Roushan2014,Yin,ZhuYu,Yao2017,Siddiqi2017,Delgado2017}. Specifically, topological concepts have been investigated in the parameter space of superconducting qubits \cite{Schroer2014,Roushan2014,Yin,ZhuYu}, including the topological Chern numbers and topological phase transitions; Topological quantum walks and Zak phases have been realized and measured in the phase space of microwave resonators \cite{Yao2017,Siddiqi2017}. Considering the-state-of-art in Xmon qubits, a natural question to ask is whether we can realize topological phases rooted in the momentum space of a superconducting Xmon qubit lattice. It is also quite interesting to study how to detect topological invariants in this qubit lattice.

In this paper, we present an experimental protocol to realize a generalized SSH model \cite{SSH} in a superconducting Xmon qubit chain with tunable qubit couplings. This model has a variety of topological magnon bands and supports different topologcial insulator phases characterized by topological winding numbers and Chern numbers. We first investigate the quantum dynamics of a single-excitation quantum state in a SSH-type qubit chain. Interestingly, we find that the time average of the center of qubit excitation difference (CED) associated with this quantum dynamics is topology-dependent and can be linked to the topological winding number. Winding number is one of the basic topological invariants but its detection method is still lacking. Our result thus gives a new method to directly measure the topological winding number using quantum dynamics of single-excitation states in the real space.

We also study the adiabatic transfer of a single-excitation quantum state in a generalized SSH-type qubit chain by slowly ramping the qubit couplings. We show that the shift of the center of  qubit excitation (CE) after one periodic ramping is exactly quantized as topological Chern number. We also find that both the amplitude and the direction of such quantized shift are entanglement-dependent. This process thus creates a novel entanglement-dependent topological pumping. Compared with topological Thouless particle pumping \cite{ThoulessPump,TPFermion,TPBoson,Mei2014,Lu}, this pumping is with respect to a single-excitation quantum state and quantum entanglement plays an important role here. This pumping can also be used to directly detect the topological Chern numbers. In contrast to recent experiments on topological properties in superconducting circuits \cite{Schroer2014,Roushan2014,Yin,ZhuYu,Yao2017,Siddiqi2017,Delgado2017}, which focus on the parameter space of the qubits and resonators, our study aims at the intrinsic topological properties associated with Bloch energy bands and rooted in momentum spaces.

This paper is organized as follows. In Sec. II, we construct a generalized SSH model with
a tunable Xmon qubit chain and study its topological features. In Sec. III, we study the time evolution of a single-excitation state in a SSH-type qubit chain. In Sec. IV, we investigate the adiabatic transfer a single-excitation state in a generalized SSH-type qubit chain by slowing ramping qubit couplings. In Sec. V, we give a summary for the main results presented in this paper.

\section{Topological states in a tunable superconducting Xmon qubit chain}
\begin{figure}[h]
\includegraphics[width=8cm,height=1.5cm]{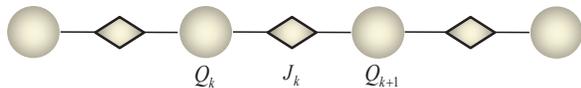}
\caption{Superconducting circuit for the one-dimensional Xmon qubit lattice. Two nearest neighbour Xmon qubits $Q_k$ and $Q_{k+1}$ are inductively coupled and the coupling strength can be tuned through the middle connected gmon coupler (CP). }
\label{Fig1}
\end{figure}

The superconducting qubit chain we consider constituts an array of coupled Xmon qubits with tunable qubit couplings, as shown in Fig. \ref{Fig1}. Suppose each Xmon qubit has two energy levels and the same transition frequency. The Hamiltonian of such Xmon qubit lattice can be descried by a spin-chain Hamiltonian
\begin{equation}
H=\sum^{L}_{k=1}J_{k}\hat{\sigma}^{\dag}_k\hat{\sigma}^{-}_{k+1}+\text{H.c}.
\end{equation}
where $\hat{\sigma}^{\dag}_k=|e\rangle_k\langle g|$, $J_{k}$ is the coupling strength between two nearest neighbour Xmon qubits. Here we omit the constant qubit frequencies and only consider singe qubit excitation, which can be precisely prepared with current superconducting qubit technology \cite{Surfacecode}. Because the number of excitations is conserved in our model, the above Hamiltonian can be reduced to single excitation subspace. Based on the Matsubara-Matsuda transformation \cite{MM}, the qubit chain can be rewritten into the following magnon Hamiltonian
\begin{equation}
H=\sum^{L}_{k=1}(J_k\hat{m}^{\dag}_k\hat{m}_{k+1}+\text{H.c}.
\label{H}
\end{equation}
where the single excitation is called as magnon in a spin chain and its annihilation operator is $\hat{m}_k=\hat{\sigma}^{-}_k$. Such Xmon qubit chain recently has been experimentally realized for studying surface-code quantum error correction \cite{QEC}. Motivated by recent experiment using gmon coupler to tune the Xmon qubit couplings \cite{Martinis2014}, we assume the coupling strength
 \begin{equation}
 J_k=g_0+g_1\cos(2\pi k/p+\theta),
  \label{Jx}
 \end{equation}
 where $g_0$ and $g_1$ are the coupling constants, $p$ is the number of qubits in one unit cell, and $\theta$ is the control parameter.

For $p=2$, each unit cell contains two qubits labeled by $a$ and $b$%
, respectively. The resulted qubit chain can be described by the SSH model Hamiltonian
\begin{equation}
\hat{H_1}=\sum_{x=1}^{N}(J_{1}\hat{a}_{x}^{\dag }\hat{b}_{x}+J_{2}\hat{b}%
_{x}^{\dag }\hat{a}_{x+1}+\text{H.c.),}  \label{SSH}
\end{equation}%
where $\hat{\alpha}_{x}^{\dag }=\hat{\sigma}_{\alpha_{x}}^{+}$ ($\alpha=a,b$) is the magnon creation operator for qubit at $%
a_{x}$ ($b_{x}$), $J_{i}=g_{0}+(-1)^{i}g_{1}\cos \theta $ ($i=1,2$) and $N$ is the
number of unit cells. To study its topological feature, we rewrite it in the momentum space as
$\hat{H}=\sum_{k_{x}}\hat{m}_{k_{x}}^{\dag }\hat{h}(k_{x})\hat{m}_{k_{x}}$, where $\hat{m}_{k_{x}}=(\hat{a}_{k_{x}},\hat{b}_{k_{x}})^{T}$, $\hat{a}_{k_{x}}$ and $\hat{b}_{k_{x}}$ are the momentum space operators,
\begin{equation}
\hat{h}(k_{x})=d_{x}\hat{\tau}_{x}+d_{y}\hat{\tau}_{y},
\label{hkssh}
\end{equation}
where $d_{x}=J_{1}+J_{2}\cos (k_{x})$, $d_{y}=J_{2}\sin (k_{x})$, and $\hat{\tau}_{x}$ and $\hat{\tau}_{y}$ are the Pauli spin operators defined in the momentum space. The energy bands of the Hamiltonian (\ref{SSH}) are characterized by topological winding number \cite{CSTopo}
\begin{equation}
\nu =\frac{1}{2\pi }\int dk_{x}\mathbf{n}\times \partial _{k_{x}}\mathbf{n}=%
\frac{1}{2}\left[ 1+\text{sgn}\left( g_{0}g_{1}\cos \theta \right) \right] ,
\label{wn}
\end{equation}%
where $\mathbf{n}=(n_{x},n_{y})=(d_{x},d_{y})/\sqrt{d_{x}^{2}+d_{y}^{2}}$. Let $g_{0}g_{1}$ be a positive number. We find that
\begin{equation}
\nu =%
\begin{cases}
1, & \mbox{$\theta \in (-\pi /2,\pi /2)$} \\
0\text{,} & \mbox{$\theta \in \left( \pi /2,3\pi /2\right)$ }%
\end{cases}%
\end{equation}%
The winding number $\nu=1$ ($0$) shows that the above SSH-type qubit chain is in the topological nontrivial (trivial) insulator phase.

For the case of $p>2$, a generalized SSH model can be formed, which supports
$p$ magnon bands. Different from the $p=2$ case, their topological features are characterized by Chern numbers. In particular, in the $p=3$ case, each unit cell
has three qubits labeled as $a$, $b$, and $c$, the Hamiltonian describing this generalized SSH-type qubit chain has the form
\begin{equation}
\hat{H_2}=\sum_{x=1}^{N}(J_{1}\hat{a}_{x}^{\dag }\hat{b}_{x}+J_{2}\hat{b}%
_{x}^{\dag }\hat{c}_{x}+J_{3}\hat{c}_{x}^{\dag }\hat{a}_{x+1}+\text{H.c.}),  \label{gSSH}
\end{equation}%
where $\hat{\alpha}_{x}^{\dag }$ ($\alpha =a,b,c$) is the magnon creation operator and $J_{s}=g_{0}+g_{1}\cos (2\pi s/3+\theta )$ ($s=1,2,3$) is the coupling strength. To explore topological features of the Hamiltonian (\ref{gSSH}), we rewrite it in the momentum space as $\hat{H}=\sum_{k_{x}}\hat{m}_{k_{x}}^{\dag }\hat{h}(k_{x},\theta )\hat{m}%
_{k_{x}}$, with $\hat{m}_{k_{x}}=(\hat{a}_{k_{x}},\hat{b}_{k_{x}},\hat{c}_{k_{x}})^{T}$,
\begin{equation}
\hat{h}(k_{x},\theta )=\sum_{i=1,4,5,6}h_{i}\hat{S}_{i},
\end{equation}
where $h_{1}=J_{1}$, $h_{4}=J_{3}\cos \left( k_{x}\right)$, $h_{5}=-J_{3}\sin \left( k_{x}\right) $,\ $h_{6}=J_{2}$, and $\hat{S}_{i}$ being the $i$-th Gell-Mann spin operator. By combining the momentum space of quasimomentum $k_{x}$ with the control variable $\theta $, we have a
synthetic two-dimensional space with parameters $\mathbf{k}=(k_{x},\theta )$%
. The energy spectrum in the first Brillouin zone $\{k_{x}\in (0,2\pi /3],$ $%
\theta \in (0,2\pi ]\}$ of this synthetic space has three energy bands. For the synthetic two-dimensional space, the underlying topological
properties of the Hamiltonian (\ref{gSSH}) are characterized by the Chern
number defined in the first Brillouin zone. Denote
the Bloch function of the $n$-th magnon band as $|u_{\mathbf{k}n}\rangle $.
The Chern number for the $n$-th magnon band is defined as \cite{TopoRev1,TopoRev2}
\begin{equation}
C_{n}=\frac{1}{2\pi }\int_{k_{x}}\int_{\theta }dk_{x}d\theta
\,F_{n}(k_{x},\theta ),  \label{CN}
\end{equation}%
where $F_{n}(k_{x},\theta )=i(|\langle \partial _{\theta }u_{\mathbf{k}%
n}|\partial _{k_{x}}u_{\mathbf{k}n}\rangle -\text{c.c.})$ is the Berry
curvature and c.c. refers to the complex conjugate. Using equation (\ref{CN}%
), we calculate the Chern numbers for the first (bottom), second (middle),
and third (top) magnon bands. The results are
\begin{equation}
\{C_{1},C_{2},C_{3}\}=%
\begin{cases}
\{2,-4,2\}, & \mbox{$-g_1/4<g_0<g_1/4$} \\
\{-1,2,1\}\text{,} & \mbox{otherwise}%
\end{cases}%
.  \label{CC}
\end{equation}%
Equation (\ref{CC}) shows that the $p=3$ generalized SSH-type qubit chain supports two
types of topological insulator phases separated by the transition points $g_{0}=\pm
g_{1}/4$.

\begin{figure}[t]
\includegraphics[width=9cm,height=8.5cm]{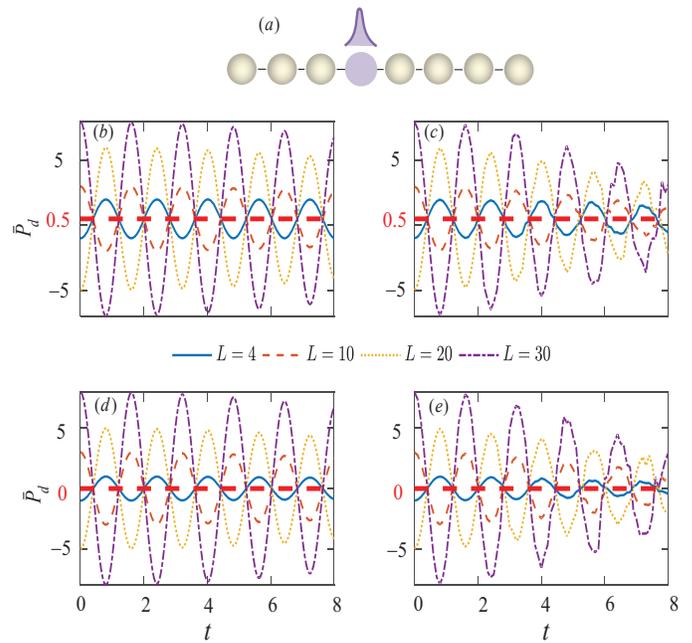}
\caption{(a) The schematic of the topology-dependent quantum
dynamics. The time-dependent  average of the CED $\bar{P}_d(t)$
versus time is shown in (a) for $\theta=0.1\pi$ with $\nu=1$ and
in (c) for $\theta=0.9\pi$ with $\nu=0$. In the presence of qubit
coupling imperfection, $\bar{P}_d(t)$ in the above two cases is
shown in (b) and (d), respectively, with the imperfect strength
$W=0.2g_1$. The other parameter is $g_0=g_1$. $g_1$ is used as
energy unit in this work.} \label{Fig2}
\end{figure}

\section{Topology-dependent quantum dynamics and winding number detection}

In this section, we will study the time evolution of a
single-excitation quantum state in  a SSH-type qubit chain (p=2).
Suppose the SSH-type qubit chain is initially prepared into a
single-excitation bulk state. As shown in Fig. \ref{Fig2}(a), we
choose to excite one of the middle qubits into the excited state
$|e\rangle$ and the other qubits are in the ground state
$|g\rangle$. Then the initial state of the system can be written
as
\begin{equation}
|\psi (0)\rangle =|gg\cdots e \cdots gg\rangle.
\end{equation}
The quantum dynamics of such single excitation state is governed by the Hamiltonian in Eq. (\ref{SSH}). After an evolution time $t$, the state of the system becomes
\begin{equation}
|\psi (t)\rangle =e^{-i\hat{H_1}t}|\psi (0)\rangle.
\end{equation}
The relation between the above quantum dynamics and the
topological feature of the SSH-type qubit  chain can be revealed
through the CED in the qubit chain. The CED is defined as
\begin{equation}
\hat{P}_{d}=\sum_{x=1}^{N}x(\hat{P}_{a_{x}}^{e}-\hat{P}_{b_{x}}^{e})
\end{equation}%
with $\hat{P}_{q}^{e}=|e\rangle _{q}\langle e|$ ($q=a_{x},b_{x}$). Then the
time-dependent average of the CED associated with the above single-excitation quantum dynamics is given by
\begin{equation}
\bar{P}_{d}(t)=\langle \psi (t)|\hat{P}_{d}|\psi(t)\rangle.
\label{pdt}
\end{equation}
Furthermore, we transfer Eq. (\ref{pdt}) into the momentum space and get
\begin{equation}
\bar{P}_{d}(t)=\frac{1}{2\pi }\int_{-\pi }^{\pi }dk_{x}\langle \psi (0)|e^{i%
\hat{h}(k_{x})t}i\partial
_{k_{x}}\hat{\tau}_{z}e^{-i\hat{h}(k_{x})t}|\psi (0)\rangle.
\label{pdtks}
\end{equation}%
By substituting Eq. (\ref{hkssh}) into Eq. (\ref{pdtks}), we find
$\bar{P}_{d}(t)$ can be connected with the  topological winding
number $\nu$ defined in Eq. (\ref{wn}), i.e.,
\begin{equation}
\begin{split}
\bar{P}_{d}(t)&=\frac{\nu }{2}-\frac{1}{4\pi}\int dk_{x}\cos(2d_{t}t)\mathbf{n}\times \partial _{k_{x}}\mathbf{n},
\label{pdk}
\end{split}
\end{equation}
where $d_{t}=\sqrt{J_1^2+J_2^2+2J_1J_2\cos(k_x)}$. The second oscillation term in Eq. (\ref{pdk}) vanishes at the critical times
\begin{equation}
t_c=(2s+1)\pi/4\sqrt{J_1^2+J_2^2},
\end{equation}
where $s$ is an integer number. At these times, the topological winding number can be directly measured via CED, i.e.,
\begin{equation}
\nu=2\bar{P}_{d}(t_c).
\label{pdtc}
\end{equation}
In the long time limit, we can also obtain a relationship between the winding number and the time-averaged CED, i.e.,
\begin{equation}
\nu ={\lim_{T\rightarrow \infty }}\frac{2}{T}\int_{0}^{T}dt\,\bar{P}_{d}(t).
\label{pdwn}
\end{equation}%
 One can find that the time-averaged CED is just the oscillation center of the CED varying with time, which dependents on the topology of the band structure of the qubit chain. Thus our result demonstrates that the quantum dynamics of a single-excitation state in a SSH-type qubit chain is topology-dependent, which can be employed for directly detecting the topological winding number.

In Figs. \ref{Fig2}(b) and \ref{Fig2}(d), we have further numerically calculated $\bar{P}_d(t)$, when the qubit chain is in the topological nontrivial and trivial phases, with the topological winding numbers $\nu=1$ and $0$, respectively. The numerical results show that $\bar{P}_d(t)$ oscillates around the average values $0.5$ and $0$, respectively, which gives the topological winding numbers $\nu=1$ and $0$ according to Eq. (\ref{pdwn}). We have also calculated $\bar{P}_d(t)$ for different choices of qubit chain lengthes. We find that the oscillation center of $\bar{P}_d(t)$ in a chain of four qubits is same as the ones in longer qubit chains. It means that the signatures of topological winding number predicted in Eq. (\ref{pdwn}) can be unambiguously observed in a qubit chain with short length.

As revealed in Eq. (\ref{pdtc}), the topological winding number can be directly detected by the CED at some critical time points. In Figs. \ref{Fig2}(b) and \ref{Fig2}(d), our numerical results show that, the oscillation curves of $\bar{P}_d(t)$ for different choices of qubit chain lengthes intersect at the time critical points $t_c$, with their values $\bar{P}_d(t_c)=0.5$ and $0$, respectively. According to Eq. (\ref{pdtc}), $\bar{P}_d(t_c)$ directly gives the topological winding number $\nu=1$ and $0$.

In practical experiments, the qubit couplings can not be perfectly tuned to exact values due to the intrinsic imperfections in device fabrication. This imperfection can be described by the Hamiltonian
$H_d=\sum_x\delta J_{1x}a^{\dag}_xb_x+\delta J_{2x}b^{\dag}_xa_{x+1}+\text{H.c.}$, where the influence of the imperfection on tuning qubit couplings is characterized by a disorder variable $\delta J_{1x,2x}=W\delta$, with $\delta\in[-0.5,0.5]$ being a random number and $W$ being the imperfect strength. In Figs. \ref{Fig2}(c) and \ref{Fig2}(e), we have taken into account the influence of qubit coupling imperfection and numerically recalculated $\bar{P}_d(t)$ for the topological nontrivial and trivial cases, respectively. For each $\delta J_{1x,2x}$, we choose 30 samples to perform the numerical simulation. The resulted CED $\bar{P}_d(t)$ is obtained by averaging over the results of all samples.
The results clearly show that $\bar{P}_d(t)$ still oscillate around $0.5$ and $0$. The critical time points where $\bar{P}_d(t)$ intersects at the oscillation center are also same as the ones shown in Figs. \ref{Fig2}(a) and \ref{Fig2}(c) without considering qubit coupling imperfection. Thus our results are quite robust to the practical imperfections in qubit coupling engineering and provide an experimentally promising method to directly detect the topological winding number.

\begin{figure}[t]
\includegraphics[width=8cm,height=7.5cm]{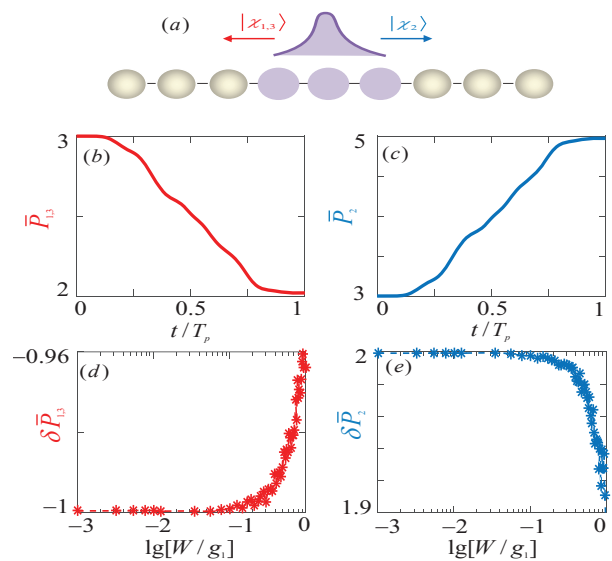}
\caption{(a) The schematic of the entanglement-dependent topological pumping. The time-dependent average of the CE $\bar{P}_n(t)$ versus time is shown in (a) and (b). The change of $\bar{P}_n(t)$ during one periodic pumping $\delta \bar{P}_n$ versus qubit coupling imperfections is shown in (c) and (d). The ramping rate
is $\Omega=0.39 g1$. The qubit number is $L=18$ and $g_0 = g_1$.}
\label{Fig3}
\end{figure}

\section{Entanglement-dependent topological pumping and Chern number detection}

In this section, we will investigate adiabatic transfer of a single-excitation quantum state in a generalized SSH-type qubit chain by slowly ramping the control parameter $\theta$. For illustration, we take $p=3$ and $g_{0}=g_{1}$, under which the three magnon bands have the Chern numbers
\begin{equation}
C_{1,3}=-1,C_{2}=2.
\end{equation}
Let the control parameter
\begin{equation}
\theta (t)=\Omega t+\varphi_{0},
\end{equation}
where $\Omega $ is the ramping rate and $\varphi _{0}$ is the initial phase. The total time for one pumping period is then $T_{p}=2\pi /\Omega$. Such time-dependent coupling has recently been implemented in superconducting Xmon qubits \cite{Roushan2016}.

At time $t=0$, let $\theta (0)=\varphi _{0}=\pi $. The coupling strengths are then $%
J_{1,2}=3g_{1}/2$ and $J_{3}=0$, i.e., the unit cells are isolated with zero
inter-cell coupling. The Hamiltonian for a single isolated unit cell in single excitation space is $\hat{H}_{s}=J_(\hat{\sigma}_{a}^{+}\hat{\sigma}_{b}^{-}+\hat{\sigma}_{b}^{+}\hat{\sigma}%
_{c}^{-})+$\text{H.c.} with $J=3g_{1}/2$. The eigenstates for such single-excitation Hamiltonian are
\begin{eqnarray}
|\chi _{1,3}\rangle &=&(|egg\rangle \mp \sqrt{2}
|geg\rangle +|gge\rangle )/2,\nonumber \\
|\chi _{2}\rangle &=&(|egg\rangle -|gge\rangle )/\sqrt{2}.
\label{chi3}
\end{eqnarray}

The corresponding eigenenergies of the above three states are $E_{1}=-\sqrt{2}J$, $%
E_{2}=0$, and $E_{3}=\sqrt{2}J$, respectively. To prepare the state $|\chi
_{1}\rangle $, we firstly decouple the selected unit cell from the rest of
the qubit chain by increasing or decreasing the detuning of qubits in this
unit cell from that of all other qubits. From the ground state $|ggg\rangle $%
, a driving pulse in the form of $\hat{V}_{3}=\Omega _{0}\cos \left( \omega _{d}t\right) \left( \hat{\sigma}_{a}^{x}-\sqrt{2}\hat{\sigma}_{b}^{x}+\hat{\sigma}_{c}^{x}\right)$
is applied with the driving frequency $\omega _{d}=\omega _{q}+\sqrt{2}J$.
In the rotating frame of $\omega _{d}$, this driving field can be written as
$\hat{V}_{3}^{\text{rot}}=\Omega _{0}(\hat{\sigma}_{a}^{x}-\sqrt{2}\hat{%
\sigma}_{b}^{x}+\hat{\sigma}_{c}^{x})/2$. It can be shown that $\langle \chi
_{1}|\hat{V}_{3}^{\text{rot}}|ggg\rangle =\Omega $ and $\langle \chi _{2,3}|%
\hat{V}_{R}|ggg\rangle =0$. By applying this pulse for a duration of $%
t_{3}=\pi /2\Omega _{0} $, the state $|\chi _{1}\rangle $ is
generated. The time duration of this operation is also on the order of nanoseconds.
Similarly, the states $|\chi _{2}\rangle $ and $|\chi _{3}\rangle $ can be
generated by applying corresponding driving pulses.

As shown in Figs. \ref{Fig3}(a), we assume the qubits in one of the
middle selected unit cells are prepared in the state $|\chi
_{n}\rangle $ $(n=1,2,3)$ defined in Eq. (\ref{chi3}) and all
other qubits are in their ground states. Then the initial state of
the qubit chain can be written as
\begin{equation}
|\psi _{n}[\theta(0)]\rangle =|ggg\cdots \chi _{n}\cdots ggg\rangle.
\end{equation}
Note that $|\psi_{n}\rangle $ is just the Wannier function of the $n$-th magnon band. When $\theta $ is swept from $t=0$ to $t=T_{p}$, the state in the initial unit cell experiences an adiabatic pumping and the entanglement will propagate to the other unit cells. Define the CE as
\begin{equation}
\hat{P}=\sum_{x=1}^{N}x(\hat{P}_{a_{x}}^{e}+\hat{P}_{b_{x}}^{e}+\hat{P}%
_{c_{x}}^{e}).
\end{equation}%
The time-dependent average of the CE for an initial excitation $|\chi _{n}\rangle $ ($n=1,2,3$) during the pumping is described by
\begin{equation}
\bar{P}_{n}(t)=\langle \psi _{n}[\theta(t)]|\hat{P}|\psi _{n}[\theta(t)]\rangle.
\label{pnt}
\end{equation}
We further write the above equation into momentum space and get
\begin{equation}
\begin{split}
\bar{P}_{n}(t)&=\frac{1}{2\pi }\int dk_{x}i\langle u_{k_{x},\theta ,n}|\partial
_{k_{x}}|u_{k_{x},\theta ,n}\rangle\\
&=\frac{1}{2\pi }\int dk_{x}A_{n}(k_{x},\theta ),  \label{PS}
\end{split}
\end{equation}%
where the Wannier function $|\psi _{n}\rangle $ has been rewritten in form
of the Bloch wave function as $|\psi _{n}\rangle =\frac{1}{2\pi }\int
dk_{x}e^{ik_{x}r}|u_{k_{x},\theta ,n}\rangle $. Equation (\ref{PS})
indicates that the CE is related to the Berry connection $%
A_{n}(k_{x},\theta )=i\langle u_{k_{x},\theta ,n}|\partial
_{k_{x}}|u_{k_{x},\theta ,n}\rangle $, which depends on the choice of the
gauge in the Bloch state. Let $\theta $ be changed continuously from $\theta
_{i}$ to $\theta _{f}$. The shift of the CE is then
\begin{equation}
\begin{split}
\bar{P}_{n}(t_{f})-\bar{P}_{n}(t_{i})&=\frac{1}{2\pi }\int
dk_{x}A_{n}(k_{x},\theta _{f})\\
&-\frac{1}{2\pi }\int dk_{x}A_{n}(k_{x},\theta_i)
\end{split}
\label{FML}
\end{equation}%
Using the Stokes theorem, equation (\ref{FML}) can be rewritten in terms of $%
F_{n}(k_{x},\theta )$ with $F_{n}(k_{x},\theta )=\nabla \times
A_{n}(k_{x},\theta )=i(|\langle \partial _{\theta }u_{\mathbf{k}n}|\partial
_{k_{x}}u_{\mathbf{k}n}\rangle -\text{c.c.})$. For a pumping circle of $2\pi
$, i.e., $\theta _{f}=\theta _{i}+2\pi $, $\hat{H}(\theta _{i})=\hat{H}%
(\theta _{f})$, and the shift of the CE is given by the
integral of the Berry curvature on the torus $\{k_{x}\in (0,2\pi /3]$,$%
\,\theta \in (0,2\pi ]\}$. We thus find
\begin{equation}
\begin{split}
\bar{P}_{n}(T_{p})-\bar{P}_{n}(0)&=\frac{1}{2\pi }\int_{k_{x}}\int_{\theta
}dk_{x}d\theta \nabla \times A_{n}(k_{x},\theta )\\
&=\frac{1}{2\pi }%
\int_{k_{x}}\int_{\theta }dk_{x}d\theta \,F_{n}(k_{x},\theta )\\
&=C_{n},
\end{split}
\end{equation}%
which shows that the shift of the CE during one pumping
circle is equal to the Chern number of the corresponding topological magnon
band and is gauge invariant.

In Figs.~\ref{Fig3}(b) and \ref{Fig3}(c), we numerically calculate $\bar{P}_{1,3}(t)$ and $
\bar{P}_{2}(t)$ for a chain of 18 qubits ($N=6$), where the qubits in the third unit cell are prepared in the state $|\chi_{1,3}\rangle$ and $|\chi_2\rangle$, respectively. It is found that
$\bar{P}_{1,3}(t)$ is shifted to the left by one unit cell and $\bar{P}_{2}(t)$ is shifted to the right by two unit cells. Moreover, the shifts of the CE are equal to the corresponding Chern numbers $C_{1,3}=-1$ and $C_{2}=2$, respectively. Such process yields an entanglement-dependent topological pumping, where both the quantized shift number and pumping direction are entanglement-dependent. This pumping can be realized with the following parameters: $g_{1}/2\pi =100$ MHz and $\Omega=0.39g_{1}$. The total pumping time is $T_{p}=2\pi /\Omega =25.5$ ns, much longer than typical decoherence times of superconducting X-mon qubits. Similarly, the entanglement-dependent topological pumping also can be realized in the generalized SSH-type qubit chain with $p>3$.

We now analyze the influence of practical imperfections in tuning qubit couplings on the above entanglement-dependent topological pumping. This imperfection can be described by the Hamiltonian
$H_d=\sum_x\delta J_{1x}\hat{a}_{x}^{\dag }\hat{b}_{x}+\delta J_{2x}\hat{b}%
_{3x}^{\dag }\hat{c}_{x}+\delta J_{3}\hat{c}_{x}^{\dag }\hat{a}_{x+1}+\text{H.c.}$, where $\delta J_{1x,2x,3x}=W\delta$, with $\delta\in[-0.5,0.5]$ being a random number and $W$ being the imperfect strength. In Figs. \ref{Fig3}(d) and \ref{Fig3}(e), we have numerically calculated
\begin{equation}
\delta \bar{P}_n=\bar{P}_n(T_p,W)-\bar{P}_n(0,W)
\end{equation}
where $\bar{P}_n(t,W)$ ($n=1,2,3$) is the CE in the presence of imperfect qubit coupling. For each
$\delta J_{1x,2x,3x}$, we choose 50 samples to perform the numerical simulation. The final derived $\bar{P}_n(t,W)$ is obtained by averaging over the results of all samples. Our numerical results show that the entangle-dependent topological pumping is robust against qubit coupling imperfections. The quantized shifts of the CE $\delta \bar{P}_{1,3}$ and $\delta \bar{P}_2$ have plateaus at the values $-1$ and $2$ when the imperfection strength $W\leq0.1g_1$, which correspond to the topological Chern numbers $C_{1,3}=-1$ and $C=2$, respectively.

\section{Conclusions}

In conclusion, we have proposed an experimentally feasible protocol using a tunable Xmon qubit chain to realize SSH and generalized SSH models that support a variety of topological magnon phases protected by topological winding numbers and Chern numbers. We have explicitly studied the dynamics of a single-excitation quantum state in these qubit chains and revealed new topological phenomenons, including the entanglement-dependent topological pumping. We have also found that the topological invariants can be directly measured from the dynamics of qubit excitation, which provides a simple method to directly measure topological invariants rooted in the momentum space using quantum dynamics in the real space. Our work may open a new prospect to realize various topological models and explore new topological effects in well-controllable quantum computing platforms.

\section{Acknowledgements}
This work is supported by the National Key R\&D Program of China under Grants No.~2017YFA0304203 and No.~2016YFA0301803; the NSFC under Grants No.~11674200, No.~11604392, No.~11434007 and No.~91636218; the PCSIRT under Grant No.~IRT13076; SFSSSP; OYTPSP; SSCC; and 1331KSC. L.T. is supported by the National Science Foundation (USA) under Award Numbers DMR-0956064 and PHY-1720501, the UC Multicampus-National
Lab Collaborative Research and Training under Award No. LFR-17-477237, and
the UC Merced Faculty Research Grants 2017.

\end{document}